\documentclass[aps,prl,twocolumn,superscriptaddress,showpacs,amsmath]{revtex4}
\usepackage{amsfonts}
\usepackage{txfonts}
\usepackage{amssymb}
\usepackage{graphicx,graphics}
\usepackage{dcolumn}
\usepackage{bm}

\begin{document}

\title{Thermal rectification and negative differential thermal
resistance in lattices with mass gradient}
\author{ Nuo Yang }

\affiliation{Department of Physics and Centre for Computational
Science and Engineering, National University of Singapore,
 117542 Singapore }

\author{ Nianbei Li }
\affiliation{Department of Physics and Centre for Computational
Science and Engineering, National University of Singapore, 117542
Singapore }

\author{ Lei Wang }
\affiliation{Department of Physics and Centre for Computational
Science and Engineering, National University of Singapore, 117542
Singapore }

\author{ Baowen Li}
\email{phylibw@nus.edu.sg }

\affiliation{Department of Physics and Centre for Computational
Science and Engineering, National University of Singapore, 117542
Singapore } \affiliation{NUS Graduate School for Integrative
Sciences and Engineering, 117597 Singapore}

\date{19 January 2007, Accepted for publication in Phys. Rev. B as Rapid Com.}
\begin{abstract}
We study thermal properties of one dimensional(1D) harmonic and
anharmonic lattices with mass gradient. It is found that the
temperature gradient can be built up in the 1D harmonic lattice
with mass gradient due to the existence of gradons. The heat flow
is asymmetric in the anharmonic lattices with mass gradient.
Moreover, in a certain temperature region the {\it negative
differential thermal resistance} is observed. Possible
applications in constructing thermal rectifier and thermal
transistor by using the graded material are discussed.
\end{abstract}

\maketitle

Theoretical studies of heat conduction in low dimensional nonlinear
lattices in past years have not only enriched our understanding of
the microscopic physical mechanism of heat conduction, but also
suggested some useful thermal device models such as
rectifier/diode \cite{diode} and thermal transistor\cite{LWC2} for
controlling heat flow. More importantly, a two segment thermal
rectifier has been realized experimentally by using
nanotubes\cite{nano}, which implies that we can control and
manipulate phonons like we do for electrons.

On the other hand, the functional graded materials (FGM) have
attracted increasing attention in many fields ranging from
aerospace, electronics, optics and bio-materials etc. due to the
intriguing properties \cite{GRIN,YuRev}. The FGM are a kind of
inhomogeneous materials whose compositions and/or structures
change gradually in space which results in corresponding changes
in physical properties such as electric, mechanical, thermal, and
optical properties. The FGM are abundant in nature, and can be
also purposely manufactured\cite{GRIN,optical}.

However, compared with optical, mechanical and many other
properties, the thermal properties of the graded materials have not
yet been fully studied (see recent review article \cite{YuRev}and
the references therein).

In this paper, we study thermal properties of the FGM represented
by 1D harmonic and anharmonic lattices with mass gradient. For
convenience, we call the lattices mass graded harmonic/anharmonic
lattices. They can be used to model superlattice or layered
structures. As it will be seen that the 1d graded lattices exhibit
new physics such as asymmetric heat flow and negative differential
thermal resistance, two essential properties used to construct
thermal rectifier and thermal transistor. Therefore, the graded
material might find new applications in heat controlling and
management.

\begin{figure}[htbp]
\centering
\includegraphics[width=9.0cm]{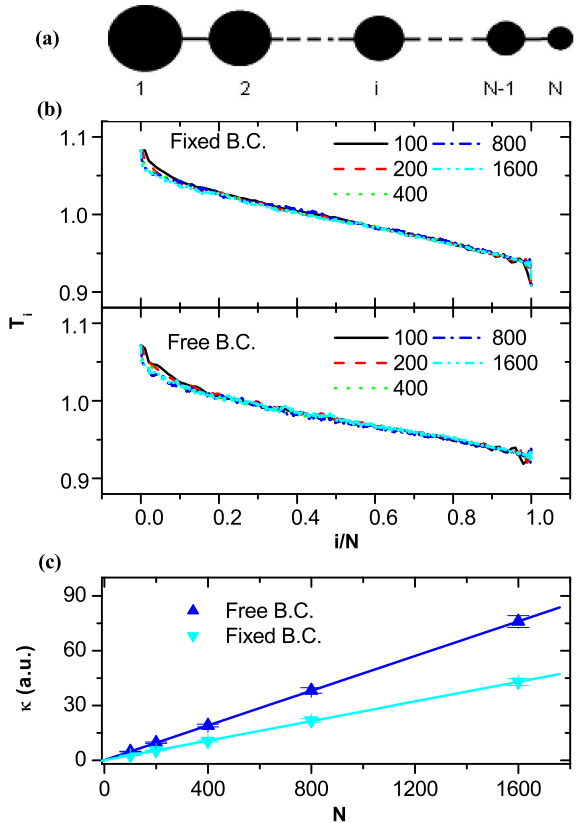}
\vspace{-1.4cm} \caption{(Color on-line) (a) Schematic picture of
the mass graded harmonic lattice.  (b) Temperature profile for the
lattice with fixed and free boundary conditions. (c) The heat
conductivity $\kappa$ versus system size $N$ for the lattice with
$ M_{MAX}=10$ and $ M_{MIN}=1$. $\kappa =AN$, where $A=0.048$ and
$0.027$ for free and fixed boundary conditions, respectively.  The
temperatures of two baths are $T_{L}=1.1$ and $T_{R}=0.9$ for left
and right end, respectively. } \label{fig:Harmonic}
\end{figure}

We consider a 1D mass graded chain, which is equivalent to a chain
with graded coupling constants\cite{Xiao}.
Fig.\ref{fig:Harmonic}(a) shows the configuration. The mass of the
$i^{th}$ particle is $ M_{i}=M_{MAX}-(i-1)(M_{MAX}-M_{MIN})/(N-1)
$, where $ M_{MAX}$ is the mass of the particle at the left end
and $ M_{MIN}$ is that of the particle at the right end. $N$ is
the total number of the particles. The Hamiltonian of this model
is:
\begin{equation}\label{ham}
H=\sum_i [\frac{p_{i}^2}{2M_i}+V(x_{i}-x_{i-1})]
\end{equation}
here $x_{i}$ is the position of the $i^{th}$ particle. Without
loss of generality, $V$ takes an anharmonic form, namely,
$(x_{i}-x_{i-1}-a)^2/2 + \beta(x_{i}-x_{i-1}-a)^4/4 $, which is a
Fermi-Pasta-Ulam (FPU) $\beta$ potential. So the system is called
a graded FPU lattice. The lattice constant $a=1.0$ and the
coupling constant $\beta=1.0$. In the case of $\beta=0$, this
lattice reduces to a graded harmonic chain.

In our simulations we use both fixed and free boundary conditions.
The stochastic heat baths are put on the $1^{st}$, $2^{nd}$,
$(N-1)^{th}$ and $N^{th}$ particles with temperature $T_{L}$ and
$T_{R}$ respectively. The equations of motion (EOM) of these four
particles are described by the Langevin equations:
\begin{equation}
M_{i} \ddot x_{i}=F(x_{i}-x_{i-1})-F(x_{i+1}-x_{i})-\left\{
\begin{array}{ll}
\xi_{L}-\lambda_{L}\dot{x}_{i},\hspace{0.3cm}&i=1,2 \\
\xi_{R}-\lambda_{R}\dot{x}_{i},\hspace{0.3cm}&i=N-1,N,
\end{array}\right.
\end{equation}
where $\xi_{L/R}$ are independent Wiener processes with zero mean,
variance $2\lambda_{L/R}k_{B}T_{L/R}$ and force $F=-\partial
V/\partial x$. To minimize the Kapitza resistance between the bath
particle and its neighbor, we set $\lambda_{L/R}/M_{i}=0.1$. Both
 the fourth-order Runge-Kutta and the velocity Verlet algorithm are
used to integrate the EOM. Differences between results of these
two integration methods are negligible. Simulations are performed
long enough ($>10^7$ time units) such that the system reaches a
stationary state where the local heat flux is constant along the
chain.

We start with the mass graded harmonic lattice, i.e. $\beta=0$ in
the potential $V(x)$. Fig.\ref{fig:Harmonic}(b) shows the
temperature profile with different lattice length $N$ and
different boundary conditions. $M_{MAX}$ and $M_{MIN}$ are fixed
at $10$ and $1$, respectively. It is well known that no
temperature gradient can be built up along a homogenous harmonic
lattice\cite{Leb}. However, in the mass graded harmonic lattice,
temperature gradient is clearly seen.

The local flux at site $i$ is defined as
\begin{equation}
J_{i}=\frac{1}{2} a \langle (\dot{x}_{i+1}+\dot{x}_{i})
F(x_{i+1}-x_{i}) \rangle.
\end{equation}
After the system reaches a stationary state, $J_{i}$ is
independent of site position $i$, so the flux can be denoted as
$J$. The heat conductivity is calculated by $\kappa =-J/(dT/dx)$.
We should stress that in calculating the temperature gradient, we
get rid of temperature jumps at the two boundaries. In
Fig.\ref{fig:Harmonic}(c), the heat conductivity versus $N$ for
different boundary conditions are plotted. For the same lattice
length $N$, heat conductivities with fixed boundaries are lower
than those with free boundaries, because there is a limitation
that all vibrational eigenmodes must vanish at boundaries in fixed
boundary cases. It is clearly seen that the heat conductivity
diverges linearly with length $N$. This linear property is
independent of boundary condition. It is different from that of
disordered harmonic lattice where the thermal conductivity
diverges with system size as $\sqrt{N}$ and $1/\sqrt{N}$ for free
and fixed boundary condition, respectively\cite{Ishii}.

The linear divergence of thermal conductivity on system size is a
new phenomenon. As we know that in 1D homogenous harmonic
lattices, no temperature gradient can be built up and thus thermal
conductivity cannot be defined\cite{Lepri1}. In order to
understand the underlying mechanism, we have to invoke the
vibrational eigenmodes of the graded system. It is found in
Ref.\cite{Xiao} that there is a critical frequency $\omega_{c}$,
which is the maximum eigenfrequency of the corresponding
homogeneous harmonic lattice with $M_{i}=M_{MAX}$, where $M_{MAX}$
is the maximal mass in the graded harmonic lattice. The modes with
$\omega < \omega_{c}$ can be well extended over the whole chain,
whereas those with $\omega
> \omega_{c}$ ( called gradons) are localized at the side with lighter
masses. Therefore, the formation of temperature gradient in the
graded harmonic lattice is attributed to the localization of the
gradons.

\begin{figure}[tbp]
\centering
\includegraphics[width=\columnwidth]{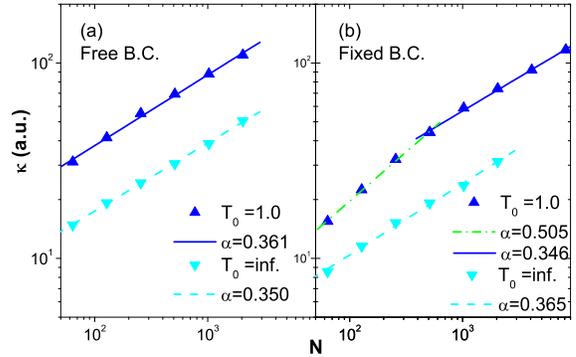}
\vspace{-1.4cm} \caption{(Color on-line) The heat conductivity
$\kappa$ versus system size $N$ for graded FPU lattice under free
boundary conditions (a) and  fixed boundary conditions (b) with $
M_{MAX}=10$ and $ M_{MIN}=1$. In all cases, $\kappa\propto
N^{\alpha}$. The values of $\alpha$ are given in the figure.
}\label{fig:FPU-KvN}
\end{figure}

In the following, we focus on the mass graded anharmonic lattice.
We first examine the size effect of heat conductivity and its
dependence on temperature. It is found that the thermal
conductivity diverges with system size as, $\kappa\sim N^\alpha$.
Fig.\ref{fig:FPU-KvN}(a) shows the temperature effect of the
divergent exponent under free boundary condition. The value of
$\alpha$ does not change very much, it almost keeps the same value
of $0.36$ ($0.35$) when the average temperature $T_{0}$ is
increased from $1.0$ to infinity. The value of the divergence
exponent is very close to the results from the renormalization
group for 1d hydrodynamic systems\cite{RG}, and that one for the
hard core model\cite{YangL}. The calculation for infinite $T_{0}$
is actually realized by discarding the quadratic term (of
potential) in Hamiltonian since this term is negligible compared
with the quartic term in infinite temperature limit.

Fig.\ref{fig:FPU-KvN}(b) shows the temperature effect of the
divergent exponent under fixed boundary conditions. In the case of
$T_0=1$, the best fitting with all available date up to $N=512$
gives rise to an $\alpha=0.51$. However, the best fitting for values
of the largest five $N$ ($512\leqslant N\leqslant 8192$) gives rise
to $\alpha=0.35$ which is very close to the value at the infinite
temperature. The $\alpha=0.51$ might be a finite size effect.

The value of $\alpha$ for free boundary condition is quite close
to the value for fixed boundary condition. Therefore, this result
seems to be very similar to the homogenous FPU-$\beta$ lattice,
namely, the divergent exponent $\alpha$ seems to be independent on
the boundary conditions.

As for the disordered FPU-$\beta$ lattice, it is observed that the
value of $\beta$ depends on the boundary condition\cite{zhaoh}.
However, this result needs to be further checked as the authors in
Ref.\cite{zhaoh} used very short lattice, $N\leq 20$.

We should point out that the above results may not be very
conclusive as we have an intermediate value of system size $N$. In
order to get a more convincing conclusion, one needs to go to
large N, say $N>10^5$, which is very difficult for current
computer facilities. However, this is not the main purpose of the
paper.

In the following, we study the thermal rectification in the mass
graded anharmonic lattice. In order to avoid the effect of the
Kapitza resistance at the boundary, we record the temperatures of
the closest particles to bath particles at two ends, namely, the
$3^{rd}$ and the $198^{th}$ for $N=200$ as $T_{l,r}$. For
convenience, we set the temperature of heavy mass end
$T_{l}=T_{0}(1+\Delta)$ and that of light mass end
$T_{r}=T_{0}(1-\Delta)$, where $T_0$ is the average temperature of
the system.

\begin{figure}[tbp]
\centering
\includegraphics[width=\columnwidth]{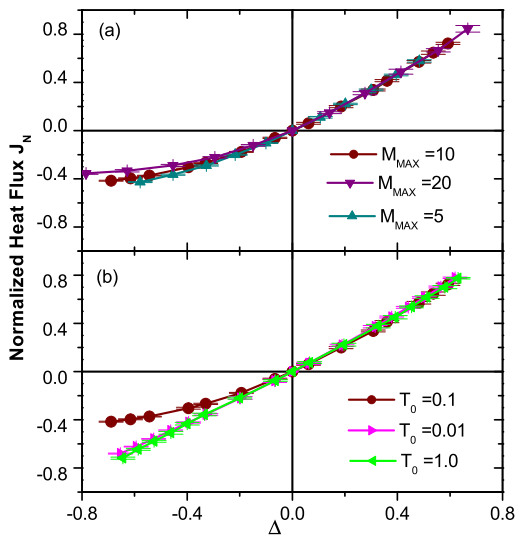}
\vspace{-1.2cm} \caption{(Color on-line) (a) Normalized Heat flux
$J_{N}$ versus $\Delta$ for three different mass gradients.
$N=200, T_{0}=0.1,M_{MIN}=1$ and $M_{MAX}=20,10$ and $5$.
$J_{N}=J/C$, where $C=0.0045, 0.0080$, and $0.0145$ for
$M_{MAX}=20,10$ and $5$ respectively. (b) Normalized Heat flux
$J_{N}$ versus $\Delta$ for $T_{0}=1.0, 0.1$, and $ 0.01$. $N=200,
M_{MIN}=1, M_{MAX}=10$. $J_{N}=J/C$, where $C=0.072, 0.008$ and
$0.0005$ for $T_{0}=1.0, 0.1$, and $0.01$ respectively.
}\label{fig:D_FPU_J}
\end{figure}

For comparison,  we show the normalized heat flux $J_{N} (\equiv
J/C)$ versus $\Delta$ in Fig.\ref{fig:D_FPU_J}(a) for three
different mass gradients, where $C$ is a constant. When $\Delta>0$
for different mass gradients, the heat flux increases steeply with
the increase of $\Delta$. However, in the case of $\Delta<0$, the
heat flux changes a little when $\Delta$ changes. This asymmetry
of heat flux with respect to $\Delta$ is called {\it thermal
rectification}. We study the dependence of the rectification on
the mass gradient (fix $M_{MIN}=1$ and  change $M_{MAX}=20,10$ and
$5$). It is shown in Fig.\ref{fig:D_FPU_J}(a) that the larger the
mass gradient, the more obvious the rectification.

To find the temperature dependence of rectification, we calculate
normalized heat flux $J_{N}$ versus $\Delta$ for different
temperature, $T_{0}=0.01, 0.1$ and $1.0$,
(Fig.\ref{fig:D_FPU_J}(b)). From this figure we can see that the
thermal rectification vanishes in both high and low temperature
limits. In the low temperature limit, the graded anharmonic
lattice reduces to the graded harmonic lattice in which no
rectification exists. In the high temperature limit, the low
frequency vibration modes, which dominate the heat conduction, in
both ends can be excited, therefore no rectification is found
either. In the case of  $T_0=0.1$, the low temperature end is
harmonic, while the high temperature end is strongly anharmonic,
thus the vibrational spectra strongly mismatch with each other,
which leads to the thermal rectification.

Finally, we should stress that the rectification is sensitive to
boundary conditions. The effect can be observed only for fixed
boundary conditions, because of the low frequency mode cannot be
restrained for free boundary conditions.

\begin{figure}[tbp]
\centering
\includegraphics[width=\columnwidth]{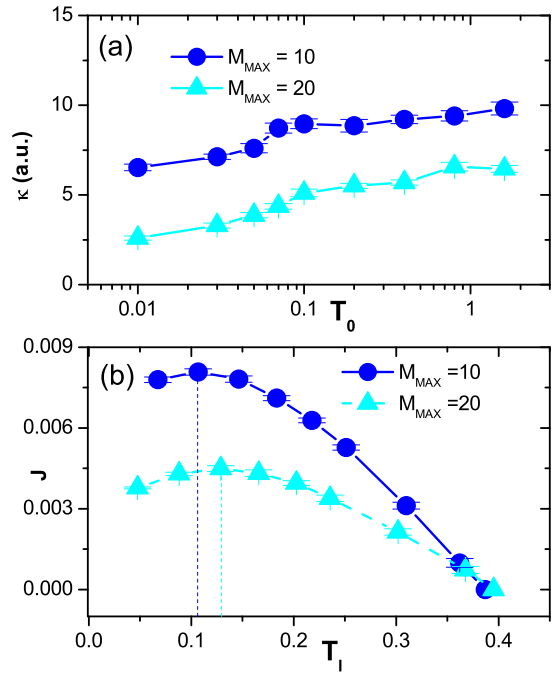}
\vspace{-1.2cm} \caption{(Color on-line) (a) Heat conductivity $\kappa$
versus temperature $T_0$. $\Delta=0.1$.
(b) Heat flux $J$ versus
$T_{l}$. $T_{r}$ is fixed at $0.4$. $N=200, M_{MIN}=1$ and
$M_{MAX}=20$ and $10$. The {\it negative differential thermal
resistances} ($R<0$) appears when $T_{l}<0.14$ and $0.11$ for
$M_{MAX}=20$ and $10$ respectively. } \label{fig:D_FPU_JvTL}
\end{figure}

We show temperature dependence of heat conductivity in
Fig.\ref{fig:D_FPU_JvTL}(a) for the graded anharmonic lattice.
Another important feature found in this 1D graded chain is the
{\it negative differential thermal resistance} (NDTR)\cite{LWC2},
namely, the larger the temperature difference the less the heat
flux through the system. In order to illustrate the NDTR, we fix
the temperature of the right (light mass) end at $T_{r}=0.4$ and
plot the heat flux $J$ versus temperature $T_{l}$ in
Fig.\ref{fig:D_FPU_JvTL}(b). The differential thermal resistance
is defined as $R=-(\partial J/\partial T_{l})^{-1}_{T_{r}=Const}$.
The NDTR is seen on the left part of the vertical line, where $J$
increases as $T_{l}$ is increased.

\begin{figure}[tbp]
\centering
\includegraphics[width=\columnwidth]{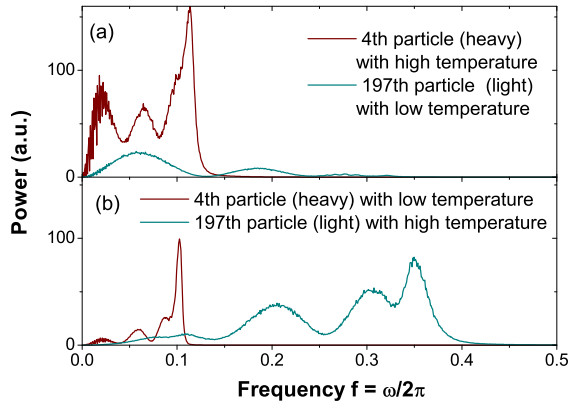}
\vspace{-1.2cm} \caption{(Color on-line) Spectra of the two end
particles for system $N=200, T_{0}=0.1, M_{MIN}=1$ and
$M_{MAX}=10$. (a) The heavy mass end (left) contact with high
temperature bath ($\Delta=0.6$) (b) The light mass end (right)
contact with high temperature bath ($\Delta=-0.7$).}
\label{fig:D_FPU_Omiga}
\end{figure}

To understand the mechanism of rectification and NDTR, we
calculate the power spectrum of particles close to the two ends,
and then compare their spectrum with each
other(Fig.\ref{fig:D_FPU_Omiga}). When temperature is high enough,
quartic term in FPU potential is the dominant term in the whole
chain. So the coupling among modes is strong and low frequency
modes, which contribute to the heat conduction most, are abundant
in the spectrum. Then the flux depends mainly on temperature
difference. So the quartic term is only dominant at the end with
high temperature and the quadratic term plays the main role at the
low temperature end, therefore, the low frequency modes cannot go
through the system. The heat current is controlled by two
competing effects: temperature gradients and overlaps of
vibrational spectra. In the presence of mass gradient, there is a
big difference between $\Delta>0$ and $\Delta<0$ in the spectra.
As shown in Fig.\ref{fig:D_FPU_Omiga}(a), where the left end
(heavy mass) contacts with high temperature bath ($\Delta>0$), the
spectra of the two particles overlap in a wide range of
frequencies, thus the heat can easily goes through the lattice
along with the direction of temperature gradient. Whereas when the
right end (light mass) contacts with high temperature bath
($\Delta<0$,Fig.\ref{fig:D_FPU_Omiga}(b)), the spectra separate
from with other. It can be seen that the light mass particle with
high temperature oscillates mainly in high frequency, however the
heavy mass particle with low temperature oscillates in low
frequency. As a result, heat is difficult to go through the system
although there is a temperature gradient. When $T_{0}$ goes to low
temperature limit, the graded chain can be regard as a graded
harmonic lattice whose thermal properties are like those of the
homogenous harmonic lattice, then the flux depends only on the
temperature difference.

In summary, we have studied thermal properties of 1D mass graded
harmonic and anharmonic lattices in this paper. It is found that
the temperature gradient can be built up in the 1D graded harmonic
lattice chain due to the localization of high frequency gradon
modes. The heat conductivity diverges with the system size
linearly, $\kappa \propto N$. In the graded anharmonic lattices,
the thermal conductivity diverges with system size as, $\kappa\sim
N^{\alpha}$, the value of $\alpha$ seems to be independent of
temperature and boundary condition.

The thermal rectification and the {\it negative differential
thermal resistance} have been observed in the graded anharmonic
lattice. This is quite similar to the recent nanotube
experiment\cite{nano}, in which half of the tube has been
gradually mass-loaded on the surface with heavy molecules. Our
results suggest that the graded materials might be used as thermal
rectifier and thermal transistor.

This work is supported in part by an Academic Research Fund from
Ministry of Education, Singapore, and the DSTA of Singapore under
Project Agreement No. POD0410553.

\end{document}